\documentclass[journal]{IEEEtran}
\usepackage{algorithmic}

\usepackage{cite}
\usepackage{color,soul}
\usepackage[rightcaption]{sidecap}

\ifCLASSINFOpdf
   
   \usepackage[pdftex]{graphicx}
   \usepackage{multirow}
   \usepackage{tabularew}
   \usepackage[flushleft]{threeparttable}

   \usepackage{algorithm}
   \usepackage{algorithmic}

   \graphicspath{{./pdf/}{./jpeg/}{./eps/}}
  \DeclareGraphicsExtensions{ ,.jpeg,.png}
\else
   \usepackage[dvips]{graphicx}
   \graphicspath{{./eps/}}
   \DeclareGraphicsExtensions{.eps}
\fi
\usepackage{amsmath}
\usepackage{array}

\ifCLASSOPTIONcompsoc
 \usepackage[caption=false,font=normalsize,labelfont=sf,textfont=sf]{subfig}
\else
  \usepackage[caption=false,font=footnotesize]{subfig}
\fi

\usepackage{stfloats}
\usepackage{url}
  \usepackage{color}

\usepackage{url}

\begin{document}

\title{X-CHANGR: Changing Memristive Crossbar Mapping for Mitigating Line-Resistance Induced Accuracy Degradation in Deep Neural Networks}
%
%
%

\author{Amogh~Agrawal, Chankyu~Lee and Kaushik~Roy,~\IEEEmembership{Fellow,~IEEE}
\IEEEspecialpapernotice{School of Electrical and Computer Engineering, Purdue University, West Lafayette, IN 47906 USA}
}
\maketitle
\pagenumbering{gobble}

\begin{abstract}

Recent advances in deep learning has accelerated the growth of machine learning and artificial intelligence in a variety of cognitive tasks. Deep learning involves a dense connection of artificial neurons and synapses to form deep neural networks (DNNs). However, DNNs are computationally and memory intensive, and consume high energy on standard von-Neumann based systems. Thus, there is widespread interest in emerging technologies, especially resistive crossbars for accelerating DNNs. Resistive crossbars offer a highly-parallel and efficient matrix-vector-multiplication (MVM) operation. MVM being the most dominant operation in DNNs makes crossbars ideally suited. However, various sources of device and circuit non-idealities lead to errors in the MVM output, thereby reducing DNN accuracy. Towards that end, we propose crossbar re-mapping strategies to mitigate line-resistance induced accuracy degradation in DNNs, without having to re-train the learned weights, unlike most prior works. Line-resistances degrade the voltage levels along the crossbar columns, thereby inducing more errors at the columns away from the drivers. We rank the DNN weights and kernels based on a sensitivity analysis, and re-arrange the columns such that the most sensitive kernels are mapped closer to the drivers, thereby minimizing the impact of errors on the overall accuracy. We propose two algorithms $-$ static remapping strategy (SRS) and dynamic remapping strategy (DRS), to optimize the crossbar re-arrangement of a pre-trained DNN. We demonstrate the benefits of our approach on a standard VGG16 network trained using CIFAR10 dataset. Our results show that SRS and DRS limit the accuracy degradation to 2.9\% and 2.1\%, respectively, compared to a 5.6\% drop from an as it is mapping of weights and kernels to crossbars. We believe this work brings an additional aspect for optimization, which can be used in tandem with existing mitigation techniques, such as in-situ compensation, technology aware training and re-training approaches, to enhance system performance.

\end{abstract}



%
\IEEEpeerreviewmaketitle

\section{Introduction}

Although artificial intelligence (AI) has been around for decades, recent advancements in deep learning (DL) has enabled machine learning and AI find value in many applications \cite{bengio2009learning,jones2014learning}. DL is based on deep neural networks (DNNs). DNNs are biologically inspired class of algorithms, which have shown state-of-the-art results for various cognitive tasks, even surpassing human intelligence in certain tasks \cite{silver2016mastering}. However, DNNs consist of a dense connection of artificial neurons and synapses, making them memory- and compute-intensive. Current computing systems are based on the well-known von-Neumann architecture, which consists of a physically separate memory and compute units. Running DNN algorithms on such machines are limited by the von-Neumann bottleneck \cite{vnbottleneck}, since the compute patterns of DNNs are inherently different. The bottleneck arises due to multiple data transfers from the off-chip memory, incurring large overheads in energy and latency. With energy efficiency being a primary concern, especially for battery operated edge devices, exploring new computing paradigms is of great importance.

In-memory computing (IMC) is one approach to overcome the von-Neumann bottleneck. IMC embeds computing within memory arrays, enabling a few computations locally where the data is stored. There have been many previous proposals for IMC for CMOS based memories, especially using SRAMs \cite{shanbhag,rsnm6t,6tml,6tddc,xsram,jiang2018xnor,xcelram,8tdpe}. However, since most DNNs are memory-intensive, having large SRAM caches that can store all weights incurs large area overheads, thereby requiring off-chip memory accesses. Embedded non-volatile memories (eNVMs), such as resistive random-access memories (ReRAM), spin-transfer-torque magnetic RAM (STT-MRAM), and phase-change materials (PCRAM), are emerging memory solutions that offer high-density storage. Moreover, the crossbar structure of such eNVMs can be leveraged to perform massively parallel matrix-vector multiplication (MVM) operations \cite{Sengupta_2016,Liu_2016,Eryilmaz_2014,Prezioso_2015}. Resistive crossbars use analog-domain for directly computing the MVM operation within the memory array itself. This makes these architectures well suited for DNNs since most of the computations in DNNs can be converted to MVM operations. Moreover, the high-density storage of eNVMs can accommodate large weights and kernels of DNNs on-chip. Multi-level resistive crossbars, which can store data into multiple conductive states, have been shown to effectively perform MVM operations for DNNs \cite{Chi_2016,Shafiee_2016,Boybat_2018,Ankit_2017,ankit2019puma,Xu_2018,trannsformer}. 


The analog nature of doing the computations in resistive crossbars induces errors and approximations in the MVM output. The sources of these errors include device and circuit non-idealities, such as device variations, line resistances, and non-idealities in the analog-digital and digital-analog converters \cite{rxnn}. These errors pose an even bigger challenge for DNNs, since the errors accumulate across deeper layers. Thus, once a trained network is mapped to the crossbars, it may not give the desired accuracy due to these errors. Many mitigation techniques have been proposed in literature to overcome these challenges, such as training on the hardware itself, or re-training the weights after being mapped onto crossbars \cite{rxnn,Chakraborty_2018,Chen_2017,fouda}. The neural network captures the error patterns and `learns' them, thereby improving the accuracy. However, these techniques require multiple writes into eNVM devices. This is a power-hungry process since eNVM writes are energy-expensive \cite{exp_mem_writes}. Moreover, low endurance of eNVM devices, especially ReRAMs, limits the number of writes into the device \cite{exp_mem_writes}. Another mitigation strategy is to lower the crossbar dimension \cite{8060457}. However, this limits the benefits of parallelism and energy-efficiency offered by crossbars. 

In this work we tackle the non-idealities through re-arranging crossbar columns, without having to re-train the learned weights. We observe a pattern in the line-resistance induced errors, which can be exploited to re-arrange the crossbar columns based on the sensitivity analysis of the weights and kernels. Line-resistances degrade the voltage levels along the crossbar columns, thereby inducing more errors at columns away from the drivers. We propose to re-map these columns based on a sensitivity analysis of the outputs. In other words, the DNN weights and kernels which are more sensitive to alter the final output are given a higher rank, and are mapped to columns closer to the drive source, thereby generating lower errors. We propose two algorithms, which take a pre-trained DNN and optimize the crossbar re-arrangement such that an improvement in the overall accuracy degradation is obtained. Note that in our work we analyze the spatial dependency among columns of the crossbar which are induced due to line-resistances. Other non-idealities like the source and sink resistances coming from peripheral circuitry affects each column equally, and do not introduce this spatial dependency. Thus, our work complements the previous efforts of mitigating crossbar non-idealities by bringing in another aspect for optimization, which can be used in tandem with existing techniques to enhance system performance.

In summary, the key highlights of this work are:
\begin{enumerate}
    
    \item We study the impact of line-resistance induced errors and spatial dependency in MVM computations in resistive crossbars, and develop a statistical model to characterize these errors.
    
    \item We propose two crossbar re-arrangement strategies - static remapping strategy (SRS) and dynamic remapping strategy (DRS). In both strategies, the crossbar arrangement of a pre-trained DNN is optimized through a sensitivity analysis of its weights and kernels.
    
    \item We evaluate the effects of line-resistance induced errors on a standard VGG16 network trained on CIFAR10 dataset, and demonstrate the improvements in accuracy degradation of the proposed mapping strategies.

\end{enumerate}

\section{Preliminaries}

In this section, we provide a brief background on resistive crossbar arrays, including their structure and operation for performing matrix-vector multiplication (MVM), and the sources of error due to parasitic line-resistances. We also briefly illustrate how large-scale DNNs are typically mapped to crossbar arrays.

\subsection{Crossbar structure and operation}

\begin{figure}[t]
\centering
\includegraphics[width=0.5\textwidth]{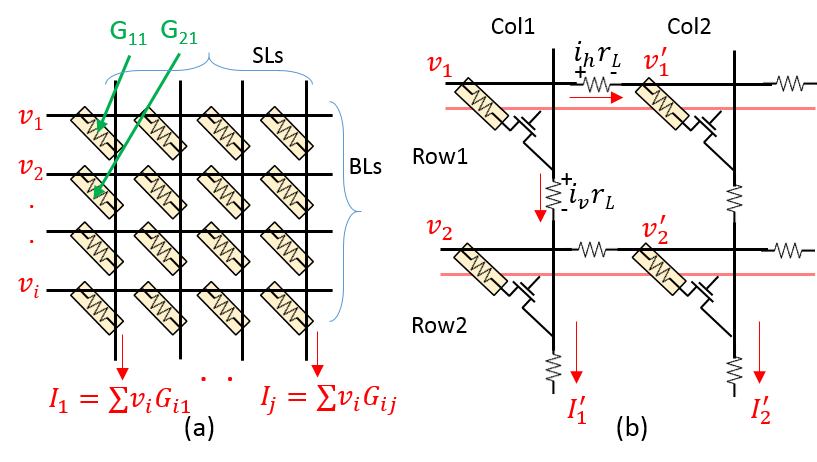}
\caption{(a) Schematic of a resistive crossbar network. The voltages ($v_i$) are applied to the horizontal BLs, and each device at the i-th row and j-th column has a conductance $G_{ij}$. The resulting current output from the SLs represent the MVM operation through Kirchoff's laws. (b) 1T-1R configuration of crossbar cells showing the voltage drops along the rows and columns. Voltages $v_1$ and $v_2$ are degraded to ${v_1}^{'}$ and ${v_2}^{'}$ due to line-resistances $r_L$.}
\centering
\label{fig:xbar}
\end{figure}

Fig. \ref{fig:xbar}(a) shows a schematic of a crossbar array. It consists of a mesh cells connected through bit-lines (BLs) running horizontally and source-lines (SLs) running vertically. Each cell is a non-volatile memory device, for example, memristor, phase-change material or a magnetic tunneling junction. Each cell also contains a selector device or a transistor, which helps read/write into individual cells and also helps block sneak-current paths \cite{xbar_selectors}. In this work, we chose a one-transistor one-resistor (1T1R) cell. To perform a matrix-vector operation, the input vector is translated to analog voltages using a digital-to-analog converter (DAC), and applied to the BLs. The matrix data is stored in the form of conductance state of the resistive elements. Each resistive element of the crossbar stores a matrix entry. The resulting current output from each SL represents the matrix-vector multiplication output obtained from Kirchoff's laws:
\begin{equation}
    I_j = \sum\limits_{i=1}^N v_iG_{ij}
\end{equation}
where $v_i$ is the analog voltage applied to i-th BL, $G_{ij}$ is the conductance of the resistive element at the crosspoint of i-th BL and j-th SL, and $I_j$ is the current output obtained at j-th SL. Thus, the crossbar structure inherently performs an MVM operation by exploiting the Kirchoff's current laws. Since most neural network computations heavily involve MVM operations, crossbars have been shown to be effective for such workloads. In that case, the input activations at each layer of the neural network are mapped to analog voltages, while the resistive devices store the learned weights of the deep neural network.

\subsection{Impact of line-resistances on crossbar operation}

Fig. \ref{fig:xbar}(b) qualitatively shows the origin of line resistance induced errors. Consider a small snippet of 2$\times$2 array of cells. Each cell, which consists of a transistor and a non-volatile memory device, has a finite size on a physical layout. Thus, the BL and SL metal lines running horizontally and vertically, respectively, have a finite resistance contribution over the length and width of the cell layout. This is depicted schematically in Fig. \ref{fig:xbar}(b), where these resistances ($r_L$) are lumped at every node of the crossbar array. First, let us consider the horizontal lines. When an input voltage is applied at the BLs, there would be voltage drops induced along the horizontal lines due to the lumped line resistances. In other words, the input voltage seen by the cells going from left to right degrades. In the example shown, $v_1$ and $v_2$ applied at Row 1 and 2 respectively, degrade to ${v_1}^{'}$ and ${v_2}^{'}$ at the second column, due to the voltage drop across $r_L$. Moreover, the amount of voltage drop at every node ($i_{h}r_{L}$) would depend on the current being drawn by that column, making it highly data-dependent on the state and the permutation of the all storage devices and input voltages in the crossbar. Next, let us consider the vertical lines. Note that as we go from the bottom to the top, the source connection of the transistor sees a higher resistance, thereby increasing the effects of source degeneration. This causes the transistor conductance to reduce, leading to errors. These errors are also data-dependent as the voltage drops along the line resistances ($i_v r_L$) depend on the current being drawn by that column. Intuitively, we get an idea that the minimum errors would be at the bottom left corner of the array, while the highest errors would be at the top right corner of the array. The data-dependency and spatial-dependency of these errors make it really difficult to estimate them quantitatively, due to large number of permutations and combinations of input voltages and the memristor states. However, by using a few key properties of DNNs, we can approximately quantify these errors, as we will show later.

\subsection{Mapping large-scale DNNs to crossbars}

\begin{figure*}[t]
\centering
\includegraphics[width=0.8\textwidth]{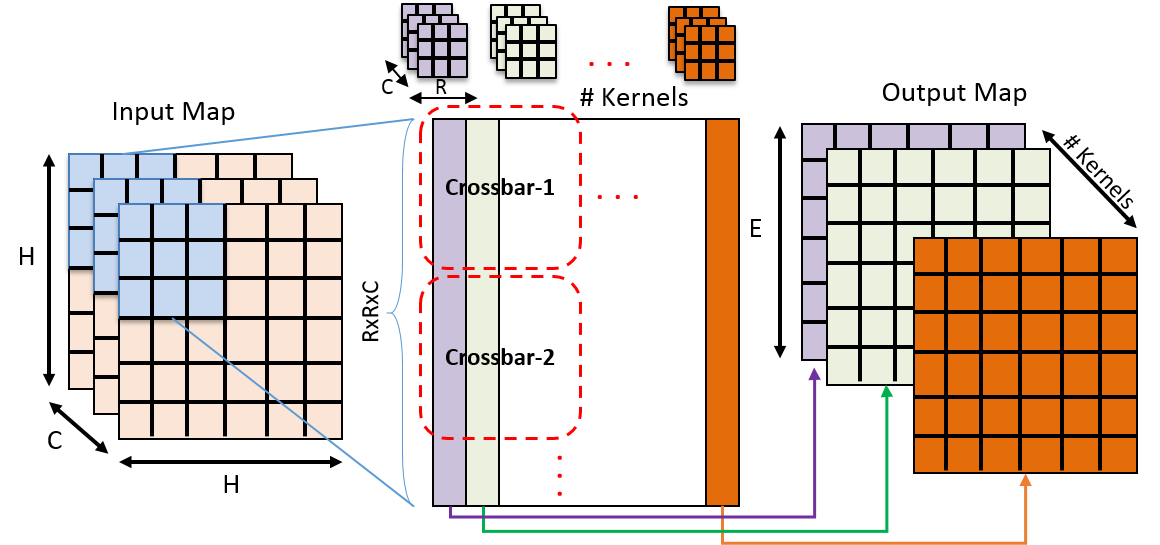}
\caption{Illustration of kernels of a convolutional neural network being mapped to crossbar arrays. Each kernel is flattened and stored as a column vector in the matrix shown. The matrix is split into multiple crossbars, where each crossbar computes a partial output, which is further accumulated to generate the output maps. Each output map is computed in a separate column of the crossbar.}
\centering
\label{fig:xbardnn}
\end{figure*}

The DNNs consist of convolutional layers (conv-) and fully-connected (fc-) layers. A conv-layer consists of multiple 3-dimensional kernels. Each kernel is flattened to a column vector and stacked, to create a big matrix. Thus, each column of the big matrix stores one kernel. The big matrix can further be divided into multiple smaller matrices corresponding to the crossbar sizes. Fig. \ref{fig:xbardnn} illustrates this process of mapping kernels to crossbars. Thus, output of each column corresponds to each output feature map. Since deeper layers of DNNs may have large number of weights, typically greater than the crossbar size, the weights are mapped to multiple crossbars where each crossbar generates a partial output. The outputs from multiple crossbars are summed to obtain the final result. Note that fc-layers can be configured as conv-layers with kernel size = input feature size, and number of kernels = number of output neurons. Thus, the proposed mapping is general to conv- and fc-layers.

In general, the weights of a DNN can have both positive and negative values. However, since the memristor conductances are positive, we use a differential architecture proposed in many previous works \cite{wpwm} to map both positive and negative weights to memristor conductances. In a differential form, each weight $w$ can be represented as $w=w^+-w^-$, where both $w^+$ and $w^-$ are positive numbers, and can be separately mapped to crossbar conductances $G^+$ and $G^-$, respectively. The output current from the positive and the negative crossbars can be subtracted to obtain the final result. Thus, Equation 1 can be written as:
\begin{equation}
\sum_{i=1}^{N} V_{i}(G_{ij}^{+}-G_{ij}^{-})= \sum_{i=1}^{N} V_{i}G_{ij}^{+}- \sum_{i=1}^N V_{i}G_{ij}^{-}=I_{j}^{+}-I_{j}^{-}
\end{equation}
where $G_{ij}^{+}$ and $G_{ij}^{-}$ are the positive and negative conductances corresponding to weight values $w_{ij}^{+}$ and $w_{ij}^{-}$, respectively, and $I_j^{+}$ and $I_j^{-}$ are the output currents from the positive and negative crossbars, respectively.

\section{Crossbar Remapping Strategies}
In this section, we propose two crossbar re-mapping algorithms, which minimize the impact of line-resistance induced errors of crossbar arrays on the system-level classification accuracy of the neural network. The idea is to map `sensitive' weights and kernels as close to the voltage drivers as possible, to have minimum output quality impact of line-resistance induced voltage drops. Thus, if all the `sensitive' weights and kernels contribute the least line-resistance induced errors in computations, the impact on final system-level classification accuracy would also be minimal.

\begin{algorithm}[!t]
\caption{SRS: Static Re-mapping Strategy}
\begin{algorithmic}[1]
 \renewcommand{\algorithmicrequire}{\textbf{Input:}}
 \renewcommand{\algorithmicensure}{\textbf{Output:}}
 
 \REQUIRE Pre-trained neural network: NN, Training example: TrData
 \ENSURE  Error Gradient: $\delta$, Neuronal Rank: $Rank$, Crossbar mapped neural network: $NN_{SRS}$ 
  \\ \textbf{\textit{Begin}}
  \STATE Initialize $NN$, $\delta_{total}$ = 0
  
  \FOR {$i = 1$ to $\# Trdata$}
   \STATE $\delta_{i}$ = Backpropagation($NN$, $Trdata_{i}$)
   \STATE $\delta_{total}$ = $\delta_{total}$ +  $\delta_{i}$
  \ENDFOR
  
  \STATE $Rank_{total}$ = EvaluateRank($\delta_{total}$)
  
  \STATE $NN_{SRS}$ = MapCrossbar($NN$, $Rank_{total}$)
  
 \RETURN $NN_{SRS}$ 
 
\end{algorithmic} 
\end{algorithm}

\subsection{SRS: Static Re-mapping Strategy}

In order to characterize the degree of sensitivity of weights and kernels to the final output quality, we use backpropagation\cite{rumelhart1985learning} technique (adopted an approach from \cite{axnn}) to calculate the (local) error gradients which is the derivatives of loss function with respect to the outputs of each neurons.
Through the backpropagation technique, one can estimate the contributions of individual neuron's output to final output error. As asserted in \cite{axnn}, the sensitive neurons contribute more to the final output error (quality) than the less sensitive ones. Thus, error gradients provide the measure of each neuron's sensitivity to impact the neural network output quality. Based on this observation, the error errors at each neuron are averaged for all instance of the training samples through backpropagation. Thus, the higher values of the accumulated error gradient are considered to be more sensitive (or important) neurons, while lower values of error gradient signifies resilient (or less important) neurons. Once we obtain local error gradients for each neuron, we rank the neurons of each layer, giving higher rank to sensitive neurons, and lower to resilient neurons.

Now that we have ranked the neurons of each layer, let us discuss how to map the weights and kernels to crossbars by utilizing the evaluated ranks. Recall from Section II-C that each column of the crossbar is mapped to weights corresponding to a particular output neuron in a fc-layer. While for conv-layers, each column is mapped to a particular kernel, which corresponds to an output feature map. For fc-layers, we directly assign crossbar columns to each output neuron based on its rank. The weights corresponding to that neuron occupy the assigned column of the crossbar. Note that the weights might span multiple crossbars, but we ensure that all weights corresponding to a particular output neuron are mapped to the same column number in all crossbars. Thus, the highest ranked neuron's weights are mapped to the first column, while the least ranked neuron's weights are mapped to the last column. For conv-layers, we take an average of the $\delta$ of all neurons corresponding to an output feature map. Next, the rankings are ascertained for each output feature map using this averaged error gradients. Since each kernel corresponds to an output feature map, the kernels are assigned crossbar columns in accordance with the ranks, similar to the fc-layer case.

We call this a static remapping strategy (SRS) since the whole analysis of calculating the ranks can be done offline, before the final mapping of conductances on crossbars. This is a one-step mapping procedure, which requires only one-time write operations to the crossbar arrays after all training examples have been evaluated.


\begin{algorithm}[!t]
\caption{DRS: Dynamic Re-mapping Strategy}
\begin{algorithmic}[1]
 \renewcommand{\algorithmicrequire}{\textbf{Input:}}
 \renewcommand{\algorithmicensure}{\textbf{Output:}}
 
 \REQUIRE  Pre-trained neural network: NN, Training example: TrData, Number of iterations: Iter
 \ENSURE  Error Gradient: $\delta$, Neuronal Rank: $Rank$, Crossbar mapped neural network: $NN_{DRS}$ 
  \\ \textbf{\textit{Begin}}
  
  \STATE Initialize $NN$
  
  \STATE $NN_{DRS}$ = $NN$, $Acc_{best}=0$
  
  \FOR {$i = 1$ to $Iter$}
   \STATE  $\delta_{i}$ = Backpropagation($NN_{DRS}$, $i$th mini-batch set of $Trdata$)
   \STATE $Rank_{i}$ = EvaluateRank($\delta_{i}$)
   \STATE  $NN_{DRS}$ = MapCrossbar($NN_{DRS}$, $Rank_{i}$)

   \STATE $Acc_{val}$ = Validation($NN_{DRS}$)
   \IF {($Acc_{val} \textgreater Acc_{best}$)}
    \STATE $Rank_{best}, Acc_{best} = Rank_{i}, Acc_{val}$
   \ENDIF
  \ENDFOR
  
 \STATE  $NN_{DRS}$ = MapCrossbar($NN_{DRS}$, $Rank_{best}$)
  
 \RETURN $NN_{DRS}$ 
 
\end{algorithmic} 
\end{algorithm}

\subsection{DRS: Dynamic Re-mapping Strategy}
We propose another re-mapping strategy by introducing the stochasticity to the SRS method. As previously mentioned, SRS is a one-step mapping strategy, where the averaged local gradients of the entire training examples were used to rank the sensitivity of neurons, and finally map the weights to the crossbar accordingly. Compared to SRS method, the dynamic remapping strategy (DRS) can be varied in terms of the number of training examples used to calculate the error gradients before mapping the weights to crossbar. In DRS, we evaluate the rank of neurons based on the mini-batches of training samples, instead of the entire training data at once. Once the ranks are evaluated for a mini-batch of training images, the weights are mapped to crossbars according to the ranks of neurons in each layer. Then, the next mini-batch of training images are used to evaluate the ranks again, and the process is repeated. Thus, the crossbars are dynamically re-mapped in this strategy. Please note that mapping strategy is analogous to the stochastic gradient descent (SGD) technique used for training neural networks with mini-batches of the training data.

However, the system-level performance ($e.g,$ classification accuracy) of the neural network does not converge while wandering the possible crossbar configurations when performing DRS method. To address this problem, we evaluate the validation accuracy using validation examples whenever re-mapping the crossbar system with updated ranks. Hence, we can store the optimal neuronal ranks of the system among large search spaces. After executing DRS method at the last set of mini-batch samples, we finally re-map the crossbar with weight sets that showed the best validation performances. As a result, the cross validation process assures to find the best neuronal rank configurations of crossbar-based neural network while iteratively searching the optimal rank of the system. 


\begin{table}[!t]
\renewcommand{\arraystretch}{1.}
\centering
\caption{Resistance ranges for various NVM technologies}
\label{table1}

\begin{tabular}{|c|c|c|}
\hline \hline
 eNVM Technology &  \bfseries $R_{ON}(\Omega)$ & \bfseries $R_{OFF}(\Omega)$\\
\hline
TaOx \cite{Kim_2016}  & 20k & 200k\\
PCM \cite{Eryilmaz_2014} & 60k & 600k\\
Ag/Si \cite{Kim_2011} & 100k & 1M\\

\hline \hline
\end{tabular}
\end{table}

\section{Evaluation Methodology}

In this section, we describe the simulation methodology that is developed to evaluate the system-level effectiveness of the proposed crossbar mapping algorithms. Firstly, we provide a detailed analysis of circuit-level crossbar modeling. Secondly, we describe the system-level simulation framework to evaluate the proposed mapping techniques on a benchmark image-recognition task using DNNs.

\subsection{Crossbar Modeling}

We use TSMC's 65nm PDK for the access transisor and an equivalent resistor to model the non-volatile memory element. The resistance values were chosen based on various memristor and phase-change material devices in literature. Different memristive technologies have different resistance ranges, from low-resistance state ($R_{ON}$) to high-resistance state ($R_{OFF}$), as shown in Table \ref{table1}. A crossbar of size 128$\times$128 was simulated in H-SPICE with each cell connected in a 1T-1R fashion. 
\begin{figure}[t]
\centering
\includegraphics[width=.5\textwidth]{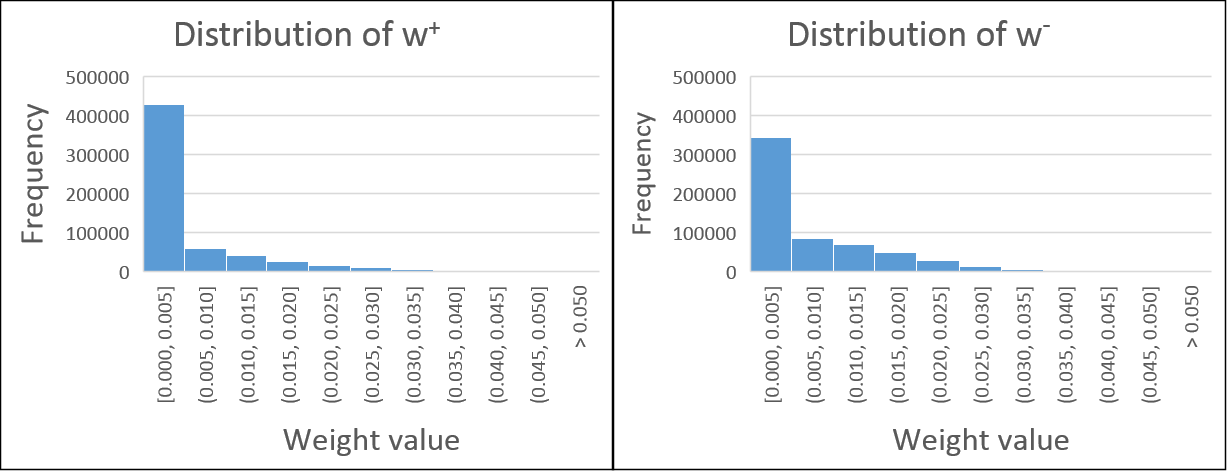}
\caption{Distribution of positive and negative weights ($w^+$ and $w^-$) for a pre-trained neural network to be mapped to crossbars. }
\centering
\label{fig:weights}
\end{figure}

We add lumped resistors along the horizontal and vertical lines at each crossbar node, to model the line-resistances. As discussed earlier, the line resistances arise due to the physical length and width of each crossbar cell, through which the BLs and SLs need to route. The layout for the 1T-1R configuration was taken from \cite{Roy_1t1r}. Since the non-volatile memory element is fabricated at the back-end-of-line (BEOL), the BLs and SLs running horizontally and vertically are usually routed on metal-2 and metal-3 layers. We used a 2$\Omega$ lumped resistor at each node, which was calculated from typical BEOL resistances and the cell area.

We use a statistical modeling approach to estimate the errors induced in the output crossbar currents due to parasitic line resistances. We saw in Section II-B that the crossbar errors increased as we go from left to right, since the line resistance induces voltage drops along the horizontal lines. We also saw that the current output at every column has a spatial-dependence on resistances of all devices in the crossbar and their respective permutations, making the analysis non-trivial and extremely difficult. In order to simplify the analysis, we employ a few key properties of DNNs. Profiling a pre-trained DNN gives us some information regarding the distribution of weights in every layer. Fig. \ref{fig:weights} plots the weight distributions for both, positive and negative crossbars, for a particular layer in the neural network. More details on the network architecture and training will be discussed in the next sub-section. We can observe, that the weights are highly skewed towards 0, which would be mapped to $R_{OFF}$, both for positive and negative crossbars. In other words, most of the devices in the crossbar array would be in $R_{OFF}$ state. We verified this assumption by taking random snippets of size 128$\times$128 from the learned kernels of the neural network, and comparing the output currents of the column-of-interest from H-SPICE, with and without replacing all other devices to $R_{OFF}$. We observed a maximum error of only $\sim$0.1\%, thereby justifying our assumption.

We randomly choose thousands of vectors $\textbf{V}$ and $\textbf{R}$ of size 128 each, from a uniform distribution $[0V,0.5V]$ and $[R_{ON},R_{OFF}]$, respectively. For each of these cases, $\textbf{R}$ was mapped to conductances of the devices in a column of the crossbar, while all other devices were kept at $R_{OFF}$. The voltages $\textbf{V}$ were applied to the BLs. The resulting current from the mapped column was recorded ($\hat{I}_j$) from H-SPICE. This was repeated for all 128 columns, by mapping $\textbf{R}$ to that column and all other devices to $R_{OFF}$, generating $\hat{I}_1$, $\hat{I}_2$,...,$\hat{I}_{128}$.

\begin{figure}[t]
\centering
\includegraphics[width=.5\textwidth]{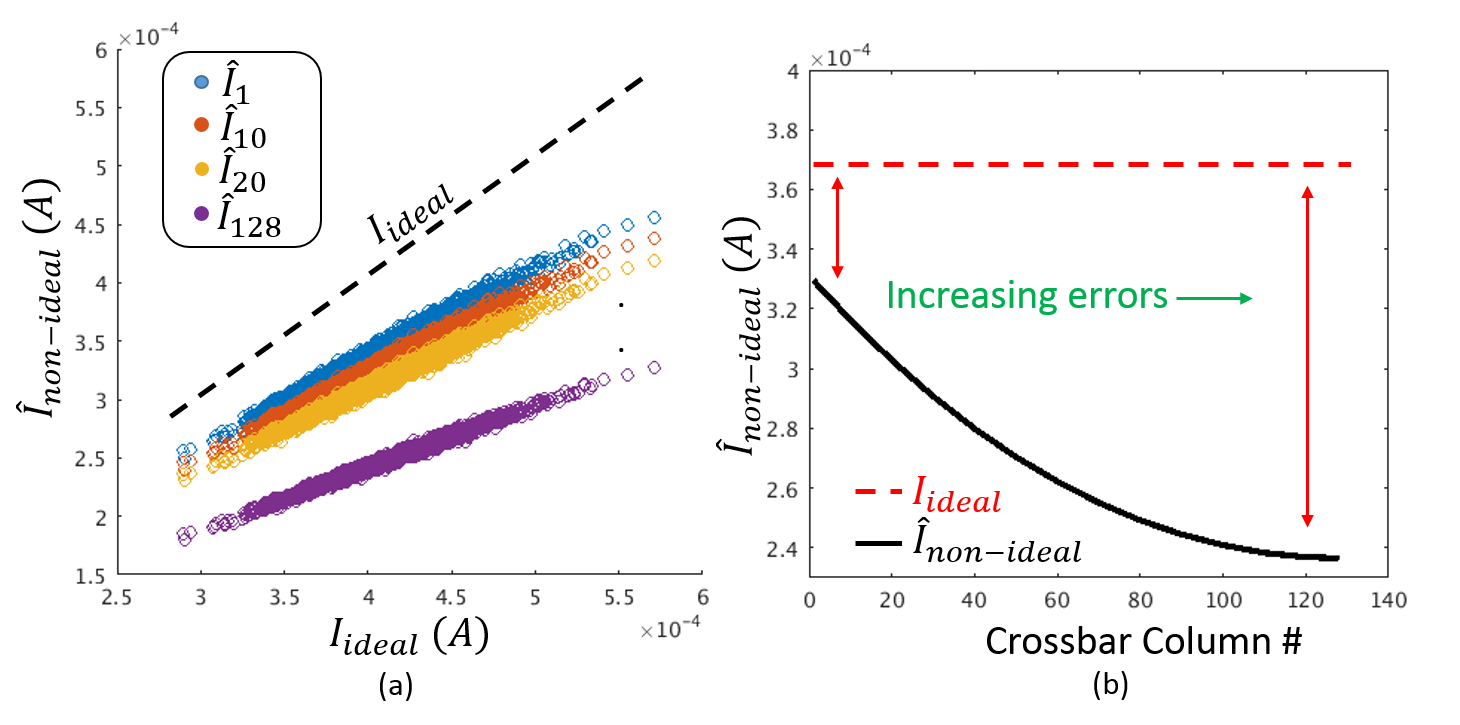}
\caption{(a) Scatter plot showing the output current from crossbar obtained from H-SPICE simulations ($\hat{I}$) as a function of the ideal expected current $I_{ideal}$, for various crossbar columns. The dotted 45$^o$ line represents the ideal case, where $\hat{I}=I_{ideal}$ (b) Picking a random current case ($I_{ideal}=370\mu A$), the figure plots $\hat{I}$ as a function of crossbar column, showing the deviation from the ideal current as the column number increases.}
\centering
\label{fig:col_errors}
\end{figure}

\begin{figure*}[t]
\centering
\includegraphics[width=\textwidth]{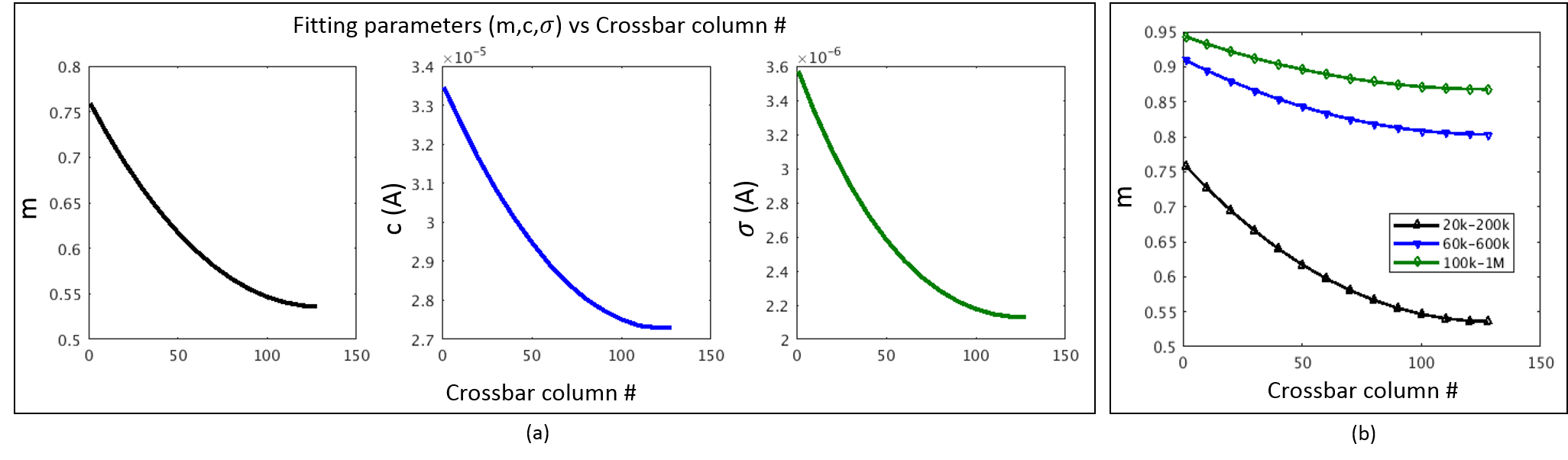}
\caption{(a) Variation of the fitting parameters $m$, $c$ and $\sigma$, as a function of crossbar column number. (b) The fitting parameter $m$ vs crossbar column number for various eNVM technologies listed in Table \ref{table1}. Other fitting parameters follow a similar trend.}
\centering
\label{fig:mcsig}
\end{figure*}

Fig. \ref{fig:col_errors}(a) shows a scatter plot, illustrating the correlation between the ideal current $I$ and the observed currents $\hat{I}_i$ from non-ideal crossbars, for various columns. Taking one random case for the current, Fig. \ref{fig:col_errors}(b) shows how the output current deviates from the ideal current as we go from the left-most column to the right-most column. A few key observations can be made from the figures. 1) At lower currents, the estimated currents closely match the ideal currents, while at higher currents, the errors are higher. This makes sense because lower currents would induce lower voltage ``$ir$'' drops along the line-resistances. 2) As the column number to which $\textbf{R}$ is mapped increases, the slope of the scatter plot increases. In other words, as we go from left-most column to right-most column, the errors in the output current increase, which is expected due to cumulative effect of line resistances. This behavior was abstracted into a crossbar model using a linear fitting:
\begin{equation}
    \hat{I}_i = m_i I + c_i + N(0,\sigma_i)
\end{equation}
where the index $i$ denotes the column number, $m$,$c$ and $\sigma$ are fitting parameters, $I$ is the ideal current output (without errors), $\hat{I}$ is the non-ideal current output, and $N$ is a normally distributed random variable with zero mean and standard deviation $\sigma$. Fig. \ref{fig:mcsig}(a) plots the fitting parameters $m$,$c$ and $\sigma$ as a function of crossbar column number. The value of $m$ drops as the crossbar column number increases, denoting the fact that the non-ideal current $\hat{I}_i$ deviates more from the ideal current $I$ as the column number increases. A similar trend is observed for the parameters $c$ and $\sigma$.

Various memristor and phase-change technologies have been proposed in literature, spanning various process techniques, materials and physics of operation. Some of these technologies have been highlighted in Table \ref{table1}, along with their $R_{ON}$ and $R_{OFF}$ values. Thus, we analyze the effects of line resistances on different $R_{ON}$ and $R_{OFF}$ values. Note that the parasitic line resistances are a function of the cell size, which is typically governed by the size of the access transistors, and the metal pitch. Assuming, the cell size remains the same for all these technologies, we expect more pronounced effects of line resistances for lower values of $R_{ON}$ and $R_{OFF}$. We repeated the above analysis for different $R_{ON}$ and $R_{OFF}$ values listed in Table \ref{table1}, and obtained the fitting parameters. Fig. \ref{fig:mcsig}(b) plots the fitting parameter $m$ as a function of crossbar column number, for various cases. It can be observed that the drop in $m$ is higher for lower resistances. This is expected, because lower the device resistances, higher the current which flows through the wires, causing larger ``$ir$'' drops. A similar trend is observed for other fitting parameters.


\subsection{System-level simulation framework}
To the analyze the effects of line-resistances at a system-level, we integrate the developed crossbar model into PyTorch deep learning framework\cite{paszke2017automatic}. We train VGG16 network\cite{simonyan2014very} using backpropagation algorithm\cite{rumelhart1985learning} on a CIFAR-10 dataset\cite{krizhevsky2009learning}. Note that we split our dataset into 3 sections (i.e. training, validation, testing) among the entire data samples. Then, we apply the proposed crossbar re-mapping algorithms (i.e. SRS and DRS) to minimize the impact of line-resistance induced errors of crossbar arrays on the system-level performance (classification accuracy). For SRS, the local gradients were averaged out on the entire training examples. Then, the weights are accordingly mapped to the crossbar depending on the evaluated ranks of neurons. For DRS, a batch-size of 8 was chosen, as it showed best results. In this case, the local gradients were averaged out after each a mini-batch iteration, to evaluate the neuronal ranks. Then, the crossbar is accordingly mapped, and the next batch is shown. This process is repeated until all training examples have been used. At each step , the validation accuracy and the ranks are recorded. We finally re-map the crossbar with optimal column ranks that showed the best validation performances.

\section{Results and Discussion}

The baseline accuracy of the trained network was observed to be 89.29\%. This is the accuracy without considering any hardware errors of the crossbars. Next, the testing data was run on the network with the developed crossbar model. In this case, the accuracy dropped to 83.70\%, a drop of 5.6\% from the baseline due to the parasitic line-resistance induced errors.

To evaluate the SRS mapping strategy, the crossbar columns were assigned to neurons based on the $\delta$'s, as described in Section III-A. The testing accuracy after the rearrangement was observed to be 86.37\%. We can clearly see an improvement in the accuracy. This is due to the fact that all the sensitive neurons, which have a higher impact on the neural network output, are mapped to crossbar columns producing the least errors. Thus, we see an overall improvement in the system accuracy. 

\begin{figure}[t]
\centering
\includegraphics[width=.35\textwidth]{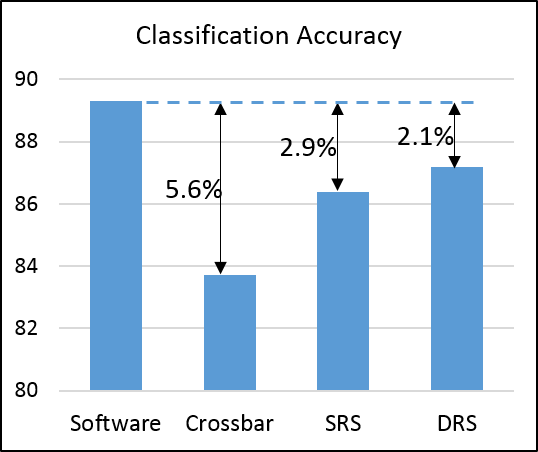}
\caption{Classification accuracy for the CIFAR10 dataset on a VGG16 network for the proposed mapping strategies $-$ SRS and DRS. The first and second bars represent the baseline accuracies, with and without crossbar errors, respectively. }
\centering
\label{fig:acc}
\end{figure}

In the DRS mapping strategy, the crossbars are re-mapped after every mini-batch of training set, and evaluated on a validation set.  The ranking scheme which gives the highest validation accuracy was saved. The best test accuracy we obtained in this case was 87.18\%. This scheme performs better than the SRS, since it involves multiple remapping steps, enabling it to explore a larger design space. Moreover, the mini-batch approach adds stochasticity, helping the system to reach different minima points. The system-level accuracy for the proposed approaches are summarized in Fig. \ref{fig:acc}.

Let us now discuss the advantages and disadvantages of the two proposed mapping strategies. Clearly, DRS method appears be the superior method than the SRS method for measuring the rank of sensitivity of neurons. However, it comes at a cost. To implement DRS, one requires multiple write-steps into the crossbar arrays. This might be unsuitable depending on the eNVM technology being used and whether the system is being deployed on a battery operated edge-device. Writing into most eNVMs are energy-expensive, and the limited endurance of eNVM devices limits the number of updates. On the other hand, SRS is an off-line approach and requires only one-time write into the crossbars.

\section{Conclusion}

Resistive crossbars have been shown to effectively accelerate DNNs, owing to their analog-domain highly-parallel MVM operation. However, various device and circuit non-idealities in crossbars induce errors in the output, which accumulates across the deeper layers. In this work, we analyzed the line-resistance induced errors in crossbars and developed a statistical model to characterize them. We proposed two algorithms to optimize the crossbar mapping, such that the effects of these line-resistances is minimized. In the first approach (SRS), we rank the weights and kernels of a pre-trained DNN using a sensitivity analysis over the entire training data-set, and assign crossbar columns according to the ranks. In the second approach (DRS), we use an iterative process of ranking and re-mapping the crossbar columns, by using mini-batches of training dataset every iteration. We integrate the statistical crossbar model into a system-level framework to analyze the accuracy degradation on a VGG16 network trained on CIFAR10 dataset. We demonstrated that the accuracy degradation was limited to only 2.9\% and 2.1\% for SRS and DRS, respectively, compared to a 5.6\% degradation an as it is mapping of weights and kernels to crossbars. We believe that our work brings in another aspect for optimization, which can be used in tandem with existing mitigation techniques to further enhance system performance.

\section*{Acknowledgements}
The research was funded in part by C-BRIC, one of six centers in JUMP, a Semiconductor Research Corporation (SRC) program sponsored by DARPA, the National Science Foundation, Intel Corporation and Vannevar Bush Faculty Fellowship. The authors would like to thank Abhishek Goyal for useful discussions on statistical modeling.

\bibliographystyle{IEEEtran}
\bibliography{sram_imc}

\begin{thebibliography}{10}
\providecommand{\url}[1]{#1}
\csname url@samestyle\endcsname
\providecommand{\newblock}{\relax}
\providecommand{\bibinfo}[2]{#2}
\providecommand{\BIBentrySTDinterwordspacing}{\spaceskip=0pt\relax}
\providecommand{\BIBentryALTinterwordstretchfactor}{4}
\providecommand{\BIBentryALTinterwordspacing}{\spaceskip=\fontdimen2\font plus
\BIBentryALTinterwordstretchfactor\fontdimen3\font minus
  \fontdimen4\font\relax}
\providecommand{\BIBforeignlanguage}[2]{{%
\expandafter\ifx\csname l@#1\endcsname\relax
\typeout{** WARNING: IEEEtran.bst: No hyphenation pattern has been}%
\typeout{** loaded for the language `#1'. Using the pattern for}%
\typeout{** the default language instead.}%
\else
\language=\csname l@#1\endcsname
\fi
#2}}
\providecommand{\BIBdecl}{\relax}
\BIBdecl

\bibitem{bengio2009learning}
Y.~Bengio \emph{et~al.}, ``Learning deep architectures for ai,''
  \emph{Foundations and trends{\textregistered} in Machine Learning}, vol.~2,
  no.~1, pp. 1--127, 2009.

\bibitem{jones2014learning}
N.~Jones, ``The learning machines,'' \emph{Nature}, vol. 505, no. 7482, p. 146,
  2014.

\bibitem{silver2016mastering}
D.~Silver \emph{et~al.}, ``Mastering the game of go with deep neural networks
  and tree search,'' \emph{Nature}, vol. 529, no. 7587, pp. 484--489, 2016.

\bibitem{vnbottleneck}
J.~Backus, ``Can programming be liberated from the von neumann style?: A
  functional style and its algebra of programs,'' \emph{Commun. ACM}, vol.~21,
  no.~8, pp. 613--641, Aug. 1978.

\bibitem{shanbhag}
M.~Kang, M.-S. Keel, N.~R. Shanbhag, S.~Eilert, and K.~Curewitz, ``An
  energy-efficient {VLSI} architecture for pattern recognition via deep
  embedding of computation in {SRAM},'' in \emph{2014 {IEEE} International
  Conference on Acoustics, Speech and Signal Processing ({ICASSP})}.\hskip 1em
  plus 0.5em minus 0.4em\relax {IEEE}, may 2014.

\bibitem{rsnm6t}
S.~Jeloka, N.~B. Akesh, D.~Sylvester, and D.~Blaauw, ``A 28 nm configurable
  memory (tcam/bcam/sram) using push-rule 6t bit cell enabling
  logic-in-memory,'' \emph{IEEE Journal of Solid-State Circuits}, vol.~51,
  no.~4, pp. 1009--1021, April 2016.

\bibitem{6tml}
J.~Zhang, Z.~Wang, and N.~Verma, ``In-memory computation of a machine-learning
  classifier in a standard 6t {SRAM} array,'' \emph{{IEEE} Journal of
  Solid-State Circuits}, vol.~52, no.~4, pp. 915--924, apr 2017.

\bibitem{6tddc}
Q.~Dong, S.~Jeloka, M.~Saligane, Y.~Kim, M.~Kawaminami, A.~Harada, S.~Miyoshi,
  D.~Blaauw, and D.~Sylvester, ``A 0.3v {VDDmin} 4+2t {SRAM} for searching and
  in-memory computing using 55nm {DDC} technology,'' in \emph{2017 Symposium on
  {VLSI} Circuits}.\hskip 1em plus 0.5em minus 0.4em\relax {IEEE}, jun 2017.

\bibitem{xsram}
A.~Agrawal, A.~Jaiswal, C.~Lee, and K.~Roy, ``{X-SRAM}: Enabling in-memory
  boolean computations in {CMOS} static random access memories,'' \emph{IEEE
  Transactions on Circuits and Systems I: Regular Papers}, pp. 1--14, 2018.

\bibitem{jiang2018xnor}
Z.~Jiang, S.~Yin, M.~Seok, and J.-s. Seo, ``Xnor-sram: In-memory computing sram
  macro for binary/ternary deep neural networks,'' in \emph{2018 IEEE Symposium
  on VLSI Technology}.\hskip 1em plus 0.5em minus 0.4em\relax IEEE, 2018, pp.
  173--174.

\bibitem{xcelram}
A.~Agrawal, A.~Jaiswal, B.~Han, G.~Srinivasan, and K.~Roy, ``Xcel-ram:
  Accelerating binary neural networks in high-throughput sram compute arrays,''
  \emph{arXiv preprint arXiv:1807.00343}, 2018.

\bibitem{8tdpe}
A.~Jaiswal, I.~Chakraborty, A.~Agrawal, and K.~Roy, ``8t {SRAM} cell as a
  multi-bit dot product engine for beyond von-neumann computing,'' \emph{arXiv
  preprint arXiv:1802.08601}, 2018.

\bibitem{Sengupta_2016}
A.~Sengupta, Y.~Shim, and K.~Roy, ``Proposal for an all-spin artificial neural
  network: Emulating neural and synaptic functionalities through domain wall
  motion in ferromagnets,'' \emph{{IEEE} Transactions on Biomedical Circuits
  and Systems}, vol.~10, no.~6, pp. 1152--1160, dec 2016.

\bibitem{Liu_2016}
C.~Liu, Q.~Yang, B.~Yan, J.~Yang, X.~Du, W.~Zhu, H.~Jiang, Q.~Wu, M.~Barnell,
  and H.~Li, ``A memristor crossbar based computing engine optimized for high
  speed and accuracy,'' in \emph{2016 {IEEE} Computer Society Annual Symposium
  on {VLSI} ({ISVLSI})}.\hskip 1em plus 0.5em minus 0.4em\relax {IEEE}, jul
  2016.

\bibitem{Eryilmaz_2014}
S.~B. Eryilmaz, D.~Kuzum, R.~Jeyasingh, S.~Kim, M.~BrightSky, C.~Lam, and
  H.-S.~P. Wong, ``Brain-like associative learning using a nanoscale
  non-volatile phase change synaptic device array,'' \emph{Frontiers in
  Neuroscience}, vol.~8, jul 2014.

\bibitem{Prezioso_2015}
M.~Prezioso, F.~Merrikh-Bayat, B.~D. Hoskins, G.~C. Adam, K.~K. Likharev, and
  D.~B. Strukov, ``Training and operation of an integrated neuromorphic network
  based on metal-oxide memristors,'' \emph{Nature}, vol. 521, no. 7550, pp.
  61--64, may 2015.

\bibitem{Chi_2016}
P.~Chi, S.~Li, C.~Xu, T.~Zhang, J.~Zhao, Y.~Liu, Y.~Wang, and Y.~Xie,
  ``{PRIME}: A novel processing-in-memory architecture for neural network
  computation in {ReRAM}-based main memory,'' in \emph{2016 {ACM}/{IEEE} 43rd
  Annual International Symposium on Computer Architecture ({ISCA})}.\hskip 1em
  plus 0.5em minus 0.4em\relax {IEEE}, jun 2016.

\bibitem{Shafiee_2016}
A.~Shafiee, A.~Nag, N.~Muralimanohar, R.~Balasubramonian, J.~P. Strachan,
  M.~Hu, R.~S. Williams, and V.~Srikumar, ``{ISAAC}: A convolutional neural
  network accelerator with in-situ analog arithmetic in crossbars,'' in
  \emph{2016 {ACM}/{IEEE} 43rd Annual International Symposium on Computer
  Architecture ({ISCA})}.\hskip 1em plus 0.5em minus 0.4em\relax {IEEE}, jun
  2016.

\bibitem{Boybat_2018}
I.~Boybat, M.~L. Gallo, S.~R. Nandakumar, T.~Moraitis, T.~Parnell, T.~Tuma,
  B.~Rajendran, Y.~Leblebici, A.~Sebastian, and E.~Eleftheriou, ``Neuromorphic
  computing with multi-memristive synapses,'' \emph{Nature Communications},
  vol.~9, no.~1, jun 2018.

\bibitem{Ankit_2017}
A.~Ankit, A.~Sengupta, P.~Panda, and K.~Roy, ``{RESPARC},'' in
  \emph{Proceedings of the 54th Annual Design Automation Conference (DAC)
  2017}.\hskip 1em plus 0.5em minus 0.4em\relax {ACM} Press, 2017.

\bibitem{ankit2019puma}
A.~Ankit, I.~E. Hajj, S.~R. Chalamalasetti, G.~Ndu, M.~Foltin, R.~S. Williams,
  P.~Faraboschi, J.~P. Strachan, K.~Roy, and D.~S. Milojicic, ``Puma: A
  programmable ultra-efficient memristor-based accelerator for machine learning
  inference,'' \emph{arXiv preprint arXiv:1901.10351}, 2019.

\bibitem{Xu_2018}
Q.~Xu, S.~Chen, B.~Yu, and F.~Wu, ``Memristive crossbar mapping for
  neuromorphic computing systems on 3d {IC},'' in \emph{Proceedings of the 2018
  on Great Lakes Symposium on VLSI (GLSVLSI)}.\hskip 1em plus 0.5em minus
  0.4em\relax {ACM} Press, 2018.

\bibitem{trannsformer}
A.~Ankit, A.~Sengupta, and K.~Roy, ``{TraNNsformer}: Neural network
  transformation for memristive crossbar based neuromorphic system design,'' in
  \emph{2017 {IEEE}/{ACM} International Conference on Computer-Aided Design
  ({ICCAD})}.\hskip 1em plus 0.5em minus 0.4em\relax {IEEE}, nov 2017.

\bibitem{rxnn}
S.~Jain, A.~Sengupta, K.~Roy, and A.~Raghunathan, ``Rx-caffe: Framework for
  evaluating and training deep neural networks on resistive crossbars,''
  \emph{arXiv preprint arXiv:1809.00072}, 2018.

\bibitem{Chakraborty_2018}
I.~Chakraborty, D.~Roy, and K.~Roy, ``Technology aware training in memristive
  neuromorphic systems for nonideal synaptic crossbars,'' \emph{{IEEE}
  Transactions on Emerging Topics in Computational Intelligence}, vol.~2,
  no.~5, pp. 335--344, oct 2018.

\bibitem{Chen_2017}
L.~Chen, J.~Li, Y.~Chen, Q.~Deng, J.~Shen, X.~Liang, and L.~Jiang,
  ``Accelerator-friendly neural-network training: Learning variations and
  defects in {RRAM} crossbar,'' in \emph{Design, Automation {\&} Test in Europe
  Conference {\&} Exhibition ({DATE}), 2017}.\hskip 1em plus 0.5em minus
  0.4em\relax {IEEE}, mar 2017.

\bibitem{fouda}
\BIBentryALTinterwordspacing
M.~E. Fouda, J.~Lee, A.~M. Eltawil, and F.~Kurdahi, ``Overcoming crossbar
  nonidealities in binary neural networks through learning,'' in
  \emph{Proceedings of the 14th IEEE/ACM International Symposium on Nanoscale
  Architectures}, ser. NANOARCH '18.\hskip 1em plus 0.5em minus 0.4em\relax New
  York, NY, USA: ACM, 2018, pp. 31--33. [Online]. Available:
  \url{http://doi.acm.org/10.1145/3232195.3232226}
\BIBentrySTDinterwordspacing

\bibitem{exp_mem_writes}
J.~{Hu}, C.~J. {Xue}, W.~{Tseng}, Y.~{He}, M.~{Qiu}, and E.~H.~. {Sha},
  ``Reducing write activities on non-volatile memories in embedded cmps via
  data migration and recomputation,'' in \emph{Design Automation Conference},
  June 2010, pp. 350--355.

\bibitem{8060457}
Y.~{Wang}, W.~{Wen}, B.~{Liu}, D.~{Chiarulli}, and H.~{Li}, ``Group scissor:
  Scaling neuromorphic computing design to large neural networks,'' in
  \emph{2017 54th ACM/EDAC/IEEE Design Automation Conference (DAC)}, June 2017,
  pp. 1--6.

\bibitem{xbar_selectors}
R.~{Aluguri} and T.~{Tseng}, ``Overview of selector devices for 3-d stackable
  cross point rram arrays,'' \emph{IEEE Journal of the Electron Devices
  Society}, vol.~4, no.~5, pp. 294--306, Sep. 2016.

\bibitem{wpwm}
M.~{Hu}, H.~{Li}, Y.~{Chen}, Q.~{Wu}, G.~S. {Rose}, and R.~W. {Linderman},
  ``Memristor crossbar-based neuromorphic computing system: A case study,''
  \emph{IEEE Transactions on Neural Networks and Learning Systems}, vol.~25,
  no.~10, pp. 1864--1878, Oct 2014.

\bibitem{rumelhart1985learning}
D.~E. Rumelhart, G.~E. Hinton, and R.~J. Williams, ``Learning internal
  representations by error propagation,'' California Univ San Diego La Jolla
  Inst for Cognitive Science, Tech. Rep., 1985.

\bibitem{axnn}
S.~{Venkataramani}, A.~{Ranjan}, K.~{Roy}, and A.~{Raghunathan}, ``Axnn:
  Energy-efficient neuromorphic systems using approximate computing,'' in
  \emph{2014 IEEE/ACM International Symposium on Low Power Electronics and
  Design (ISLPED)}, Aug 2014, pp. 27--32.

\bibitem{Kim_2016}
K.~M. Kim, J.~J. Yang, J.~P. Strachan, E.~M. Grafals, N.~Ge, N.~D. Melendez,
  Z.~Li, and R.~S. Williams, ``Voltage divider effect for the improvement of
  variability and endurance of {TaOx} memristor,'' \emph{Scientific Reports},
  vol.~6, no.~1, feb 2016.

\bibitem{Kim_2011}
K.-H. Kim, S.~Gaba, D.~Wheeler, J.~M. Cruz-Albrecht, T.~Hussain, N.~Srinivasa,
  and W.~Lu, ``A functional hybrid memristor crossbar-array/{CMOS} system for
  data storage and neuromorphic applications,'' \emph{Nano Letters}, vol.~12,
  no.~1, pp. 389--395, dec 2011.

\bibitem{Roy_1t1r}
J.~{Li}, , S.~{Salahuddin}, and K.~{Roy}, ``Variation-tolerant spin-torque
  transfer (stt) mram array for yield enhancement,'' in \emph{2008 IEEE Custom
  Integrated Circuits Conference}, Sep. 2008, pp. 193--196.

\bibitem{paszke2017automatic}
A.~Paszke, S.~Gross, S.~Chintala, G.~Chanan, E.~Yang, Z.~DeVito, Z.~Lin,
  A.~Desmaison, L.~Antiga, and A.~Lerer, ``Automatic differentiation in
  pytorch,'' 2017.

\bibitem{simonyan2014very}
K.~Simonyan and A.~Zisserman, ``Very deep convolutional networks for
  large-scale image recognition,'' \emph{arXiv preprint arXiv:1409.1556}, 2014.

\bibitem{krizhevsky2009learning}
A.~Krizhevsky and G.~Hinton, ``Learning multiple layers of features from tiny
  images,'' Citeseer, Tech. Rep., 2009.

\end{thebibliography}

\end{document}